\documentclass[aps,prb,twocolumn,showpacs,floatfix]{revtex4-1}

  \usepackage[utf8]{inputenc}
  \usepackage[T1]{fontenc}
  \usepackage{geometry}
  \usepackage{float} 
  \usepackage{color}  
\usepackage{graphicx}
\usepackage{amsmath}
\usepackage{bbm}
\usepackage{indentfirst}
\usepackage{dcolumn}
\usepackage{soul}

\begin{document}

\title{Magnetic phase diagram of an Fe monolayer on W(110) and Ta(110) surfaces based on \textit{ab initio} calculations}

\author{Levente R\'ozsa}
\email{rozsa@phy.bme.hu}
\affiliation{Department of Theoretical Physics, Budapest University of Technology and Economics, Budafoki \'ut 8, H-1111 Budapest, Hungary}
\author{L\'aszl\'o Udvardi}
\author{L\'aszl\'o Szunyogh}
\affiliation{Department of Theoretical Physics, Budapest University of Technology and Economics, Budafoki \'ut 8, H-1111 Budapest, Hungary\\ MTA-BME Condensed Matter Research Group, Budapest University of Technology and Economics, Budafoki \'ut 8, H-1111 Budapest, Hungary}
\author{Istv\'{a}n A. Szab\'{o}}
\affiliation{Department of Solid State Physics, University of Debrecen, H-4010 Debrecen P.O.Box 105, Hungary}
\date{\today}

\begin{abstract}

We present detailed investigations of the magnetic properties of an Fe monolayer on W and Ta $(110)$ surfaces based on the \textit{ab initio} screened Korringa--Kohn--Rostoker method. By calculating tensorial exchange coupling coefficients, the ground states of the systems are determined using atomistic spin dynamics simulations. Different types of ground states are found in the systems as a function of relaxation of the Fe layer. In case of W$(110)$ substrate this is reflected in a reorientation of the easy axis from in-plane to out-of-plane. For Ta$(110)$ a switching appears from the ferromagnetic state to a cycloidal spin spiral state, then to another spin spiral state with a larger wave vector and, for large relaxations, a rotation of the normal vector of the spin spiral is found. Classical Monte Carlo simulations indicate temperature-induced transitions between the different magnetic phases observed in the Fe/Ta$(110)$ system. These phase transitions are analyzed both quantitatively and qualitatively by finite-temperature spin wave theory.

\end{abstract}

\maketitle

\section{Introduction}

The Dzyaloshinsky--Moriya interaction\cite{Dzyaloshinsky,Moriya} between local magnetic moments has a great impact in spintronics applications through the formation of chiral spin structures like magnetic skyrmions\cite{Heinze,Romming} and chiral domain walls,\cite{Chen} while it may also lead to an asymmetry in the magnon spectrum of ferromagnetic thin films, as was shown theoretically\cite{Udvardi} and examined in spin-polarized electron energy loss spectroscopy experiments\cite{Zakeri,Zakeri2} for Fe/W$(110)$. Spin-polarized scanning tunneling microscopy experiments enabled the real-space observation of spin spiral orderings at low temperatures in several ultrathin films such as Mn monolayer on W$(110)$,\cite{Bode} Pd/Fe double-layer on Ir$(111)$,\cite{Romming} Cr monolayer on W$(110)$\cite{Santos} and Fe double-layer on W$(110)$.\cite{Kubetzka,Meckler} 

A double-layer of Fe on W$(110)$ shows unusual phase transitions when the temperature is increased. While the monolayer is ferromagnetic up to $T_{c}\approx 230\,\textrm{K}$,\cite{Elmers} in the double-layer the low-temperature spin spiral phase disappears at around $200\,\textrm{K}$,\cite{Bergmann} developing an in-plane ferromagnetic state as in the case of the monolayer, which persists up to $T_{c}\approx 450\,\textrm{K}$.\cite{Elmers} This is in agreement with the asymmetry of the spin wave spectrum found in Ref. [\onlinecite{Zakeri}] at $T\approx 300\,\textrm{K}$, since the spectrum around a cycloidal spin spiral ground state would be symmetric if the Dzyaloshinsky--Moriya interaction were perpendicular to the plane of the spiral.\cite{Michael,Michael2} Using the experimentally obtained wavelength of the low-temperature spiral state it was possible to find micromagnetic exchange (spin stiffness), Dzyaloshinsky--Moriya and anisotropy parameters describing this type of order.\cite{Meckler,Meckler2} However, both micromagnetic\cite{Heide,Zimmermann} and atomistic\cite{Bergqvist} \textit{ab initio} calculations indicated a ferromagnetic ground state in the system. For an Fe monolayer on W$(110)$, theoretical calculations\cite{Qian,Nakamura,Bergman} agree with experiments\cite{Elmers2} in determining an in-plane ferromagnetic ground state. For a Mn monolayer on W$(110)$, Ref.~[\onlinecite{Bode}] provided consistent experimental and theoretical descriptions of the spiral ground state. \textit{Ab initio} calculations\cite{Zimmermann} and experiments\cite{Santos} are also in agreement about the spiral ground state of Cr monolayer on W$(110)$.

Various types of magnetic ground state configurations were found by \textit{ab initio} calculations in an Fe monolayer on the $(100)$ surface of W$_{1-x}$Ta$_x$ ($0 \le x \le 1$) alloys\cite{Ferriani,Ondracek} as a function of Ta concentration $x$, ranging from an antiferromagnetic state on pure W to a ferromagnetic state on pure Ta. Both W and Ta have bcc lattice structure but the lattice constant of Ta is about $4.3\%$ larger than that of W  ($a_{\rm Ta}=3.301\,$ \AA\ and $a_{\rm W}=3.165\,$ \AA\ ). This difference was taken into account by calculating the lattice constant of the alloy, but the relaxation of the Fe layer towards the top substrate layer was kept fixed during the calculations at the value determined for Fe/W$(100)$, although Fe should have a larger inward relaxation in the case of Ta with the larger lattice constant. For different relaxations, Fe on Ta$(100)$ may have either ferromagnetic or antiferromagnetic ground state as shown in Ref.~[\onlinecite{Simon}].

In this paper we examine the magnetic ground state of an Fe monolayer on W and Ta $(110)$ surfaces as a function of the relaxation of the Fe layer with respect to the top substrate layer. The electronic structure calculations were performed by using the relativistic screened Korringa-Kohn-Rostoker method.\cite{Szunyogh2} For the determination of the magnetic ground state we mapped the spin system onto a generalized classical Heisenberg model, where the parameters are taken from the relativistic generalization\cite{Udvardi2} of the method of infinitesimal rotations introduced by Liechtenstein \textit{et al.}\cite{Liechtenstein} The ground state of the system was found by atomistic spin dynamics simulations based on the Landau-Lifshitz-Gilbert\cite{Landau,Gilbert} equations. These results are described in Sec.~\ref{sec2}.

Besides changing the relaxation, thermal fluctuations may also induce transitions between the different types of ordered states found in these systems. Classical Monte Carlo simulations were performed using the previously obtained spin model to find these transitions. In one of the transitions found in Fe monolayer on Ta$(110)$ the increasing temperature drives the system from the ferromagnetic ground state into a non-collinear spin spiral state. Most likely, this transition is driven by the Dzyaloshinsky--Moriya interactions and the easy-axis anisotropy in the system. Such a transition was already studied in Refs. [\onlinecite{Dzyaloshinsky2}] and [\onlinecite{Izyumov}] using a Ginzburg--Landau model, which is, however, unsuitable for employing Heisenberg model parameters obtained from \textit{ab initio} calculations.

Instead of relying on a continuum model, we used spin wave expansion to describe the transition between the different ordered states. This method was found to be a powerful tool\cite{Yosida,Miwa} for explaining a transition from a low-temperature ferromagnetic to a high-temperature helical state in bulk Dy. In the present work we incorporated the Dzyaloshinsky--Moriya interaction into such an analysis, which was unnecessary in bulk systems with an inversion center, but it plays an important role in case of ultrathin films. We also used the spin wave expansion technique to handle higher order terms (magnon-magnon interactions) perturbatively, since perturbation theory makes it possible to estimate the temperature where the system reaches the paramagnetic state. This method was originally used to calculate the Curie temperature in a simple cubic lattice described by a ferromagnetic Heisenberg model.\cite{Bloch} By using a simplified model Hamiltonian consistent with the different types of ground states found in an Fe monolayer on Ta$(110)$, in Sec.~\ref{sec3} we present a detailed analysis of the temperature-induced magnetic phase transitions and relate the results to those obtained from Monte Carlo simulations.

\section{Magnetic states and phase transitions in an Fe monolayer on W and Ta $(110)$ surfaces \label{sec2}}

\subsection{\textit{Ab initio} calculation of collinear magnetic states\label{sec2A}}

For the \textit{ab initio} calculations we used the relativistic screened Korringa-Kohn-Rostoker method,\cite{Szunyogh,Zeller,Szunyogh2} using the local spin density approximation and the atomic sphere approximation. First we performed calculations for W and Ta bulk with the lattice constants $a_{\textrm{W}}=3.165$\,\AA\, and  $a_{\textrm{Ta}}=3.301$\,\AA, respectively. The layered systems considered for the deposited Fe monolayers  
comprised eight layers of bulk atoms, one layer of Fe and three layers of empty spheres, 
sandwiched between the semi-infinite bulk calculated in the previous step and a semi-infinite vacuum. Theoretical calculations using the full-potential linearized augmented plane-wave method give relaxation values between $12-13\%$ for an Fe monolayer on W$(110)$,\cite{Qian2,Qian3,Bergman,Huang} while the experimental values are in the range of $7-13\%$.\cite{Albrecht,Tober,Meyerheim} On Ta$(110)$ Fe should have an even larger relaxation due to the larger lattice constant. Therefore the calculations were performed for different values of the distance between the Fe monolayer and the top bulk monolayer, adjusting the Wigner--Seitz radius of the atomic spheres related to the Fe atoms correspondingly. Both for W and Ta, the relative relaxation with respect to the ideal distance between bcc$(110)$ atomic layers was changed between $10\%$ and $17\%$. All the atomic layers but the Fe layer were kept at the ideal lattice geometry since calculations\cite{Qian2,Qian3,Huang} indicate that the W-W relaxations are below $1\%$ even between the topmost W monolayers. We determined the potential and the exchange-correlation magnetic field self-consistently, serving as an input to the evaluation of the exchange coefficients, see Sec.~\ref{sec2B}.

The spin and orbital magnetic moments obtained from the \textit{ab initio} calculations are listed in Table \ref{table1}. The sum of the spin and orbital moments in the Fe layer on W(110) for $13\%$ inward relaxation compares within $10\%$ to the total magnetic moments given in the literature.\cite{Qian2,Qian3,Huang} The induced moments in the topmost W layer are antiparallel to the Fe moments, in agreement with Refs.~[\onlinecite{Qian2}] and [\onlinecite{Qian3}], but they are parallel in the next two W layers. It is worth noting that the spin and orbital moments are parallel for the W atoms although the W $d$--shell is less than half-filled, which indicates a violation of Hund's third rule, cf. Ref.~[\onlinecite{Qian3}]. Apparently, this is not the case for Ta. It is also notable that the induced moments of the Ta atoms are larger than those of the corresponding W atoms. Ref.~[\onlinecite{Huang}] agrees with our calculation inasmuch as increasing the relaxation decreases the magnetic moments of the Fe atoms, most likely due to the increased hybridisation between the Fe and the substrate layers. 

\begin{table*}
\begin{ruledtabular}
\begin{tabular}{ccccccccc}

           &                                                                        \multicolumn{ 8}{c}{Fe/W(110)} \\

           &                  \multicolumn{ 4}{c}{spin moment ($\mu_{\textrm{B}}$)} &               \multicolumn{ 4}{c}{orbital moment ($\mu_{\textrm{B}}$)} \\

relaxation &        Fe1 &         W1 &         W2 &         W3 &        Fe1 &         W1 &         W2 &         W3 \\

      10\% &      2.355 &     -0.164 &      0.007 &      0.003 &      0.180 &     -0.027 &      0.001 &     -0.001 \\

      13\% &      2.244 &     -0.164 &      0.012 &      0.003 &      0.169 &     -0.018 &      0.005 &      0.000 \\

      15\% &      2.181 &     -0.161 &      0.017 &      0.004 &      0.162 &     -0.014 &      0.007 &      0.001 \\

      17\% &      2.122 &     -0.156 &      0.022 &      0.004 &      0.156 &     -0.011 &      0.010 &      0.002 \\ \\

           &                                                                       \multicolumn{ 8}{c}{Fe/Ta(110)} \\

           &                  \multicolumn{ 4}{c}{spin moment ($\mu_{\textrm{B}}$)} &               \multicolumn{ 4}{c}{orbital moment ($\mu_{\textrm{B}}$)} \\

relaxation &        Fe1 &        Ta1 &        Ta2 &        Ta3 &        Fe1 &        Ta1 &        Ta2 &        Ta3 \\

      10\% &      2.587 &     -0.278 &     -0.037 &     -0.027 &      0.100 &      0.031 &      0.005 &      0.003 \\

      13\% &      2.520 &     -0.310 &     -0.045 &     -0.030 &      0.097 &      0.036 &      0.006 &      0.003 \\

      15\% &      2.466 &     -0.333 &     -0.046 &     -0.027 &      0.094 &      0.040 &      0.006 &      0.002 \\

      17\% &      2.406 &     -0.358 &     -0.044 &     -0.023 &      0.090 &      0.044 &      0.006 &      0.001 \\

\end{tabular}  
\end{ruledtabular}
\caption{Calculated spin and orbital moments in the Fe layer and in the topmost three substrate layers of W(110) and Ta(110) surfaces for selected values of relaxations of the Fe layer.}
\label{table1}
\end{table*}

\subsection{Calculated exchange interactions\label{sec2B}}

Using the self-consistent potentials obtained before, the relativistic torque method\cite{Udvardi2} was employed to map the energy of the magnetic system onto a generalized Heisenberg model,
\begin{equation}
H=\frac{1}{2}\sum_{\substack {i,j \\  (i \ne j)} } J_{ij}^{\alpha\beta}S_{i}^{\alpha}S_{j}^{\beta}+\sum_{i}K_{i}^{\alpha\beta}S_{i}^{\alpha}S_{i}^{\beta},\label{eqn1}
\end{equation}
where $i,j$ and  $\alpha,\beta$ label lattice sites and Cartesian indices, respectively,  ${S}^\alpha_{i}$ are the components of the unit vector representing the orientation of the spin at lattice site $i$, while $J_{ij}^{\alpha\beta}$ and $K_{i}^{\alpha\beta}$ stand for the matrix elements of the exchange coupling tensors and of the second-order on-site anisotropy energy tensors. The relativistic torque method relies on the magnetic force theorem and requires the calculation of coupling coefficients around different collinear reference states for at least three linearly independent magnetization directions, since for a given direction, only those components of the $\boldsymbol{J}_{ij}$ tensors can be obtained which lie in the plane perpendicular to the magnetization. In particular, we considered the magnetization directions $[1\overline{1}0]$, $[001]$, and $[110]$. The spins in a given layer must be ferromagnetically aligned, but the antiferromagnetic ordering between the different layers was taken into account. To perform the necessary integrations, $16$ energy points were taken along a semicircle contour in the upper complex semiplane, and from $204$ up to $6653$ $\boldsymbol{k}$-points were sampled in the Brillouin zone, gradually increasing for energies approaching the Fermi level.

\begin{figure}
\includegraphics[width=\columnwidth]{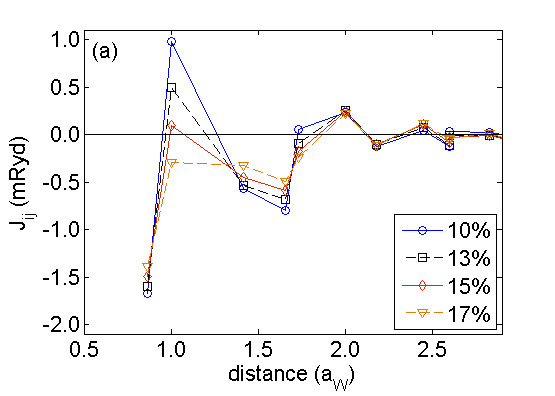}
\includegraphics[width=\columnwidth]{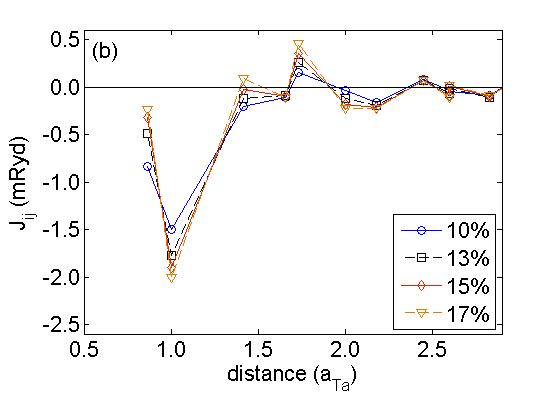}
\caption{(color online) Calculated isotropic exchange interactions $J_{ij}$ obtained from the relativistic torque method, for (a) W(110) and (b) Ta(110) surfaces and different values of relaxations of the Fe layer.}
\label{figJij}
\end{figure}

\begin{figure*}
\includegraphics[width=\columnwidth]{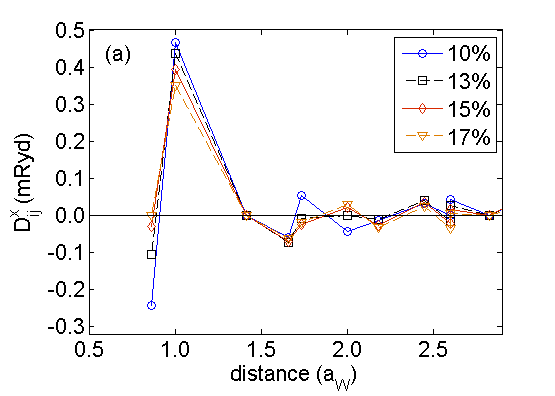}
\includegraphics[width=\columnwidth]{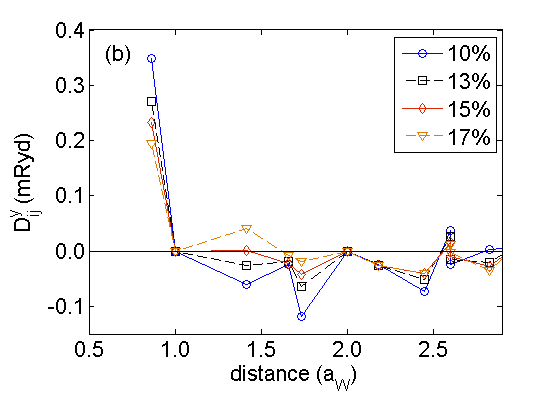}
\includegraphics[width=\columnwidth]{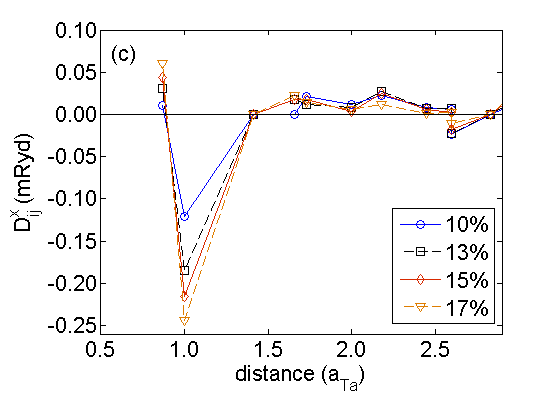}
\includegraphics[width=\columnwidth]{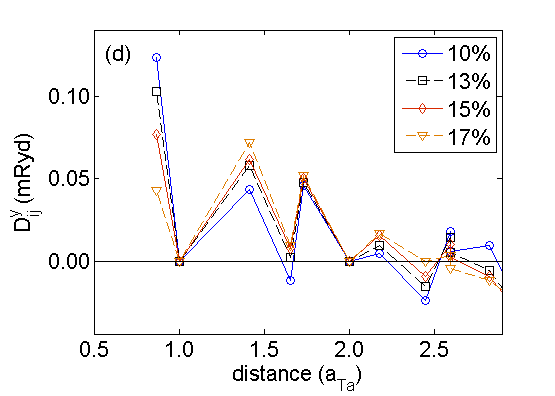}
\caption{(color online) Calculated components of the Dzyaloshinsky-Moriya vectors, $D_{ij}^{x}$ and $D_{ij}^{y}$, in an Fe monolayer on (a)-(b) W(110) and (c)-(d) Ta(110) surfaces for different values of relaxations of the Fe layer. Displayed are the values for the pairs with $R_{ij}^{x}>0$ and $R_{ij}^{y}>0$, while
the Dzyaloshinsky-Moriya vectors for the rest of the pairs can be obtained from the symmetry relations described in the text.}
\label{figDij}
\end{figure*}

The isotropic part of the exchange tensors between the Fe atoms,
\begin{eqnarray}
J_{ij}=\frac{1}{3}\sum_{\alpha}J_{ij}^{\alpha\alpha},
\end{eqnarray}
is shown in Fig.~\ref{figJij}, for W and Ta surfaces and different relaxation values. Note that with the sign convention of Eq. (\ref{eqn1}), $J_{ij}<0$ and $J_{ij}>0$ indicate ferromagnetic and antiferromagnetic couplings, respectively. In case of W(110) surface, the nearest-neighbor ferromagnetic coupling is fairly insensitive to the relaxation, while the next-nearest-neighbor coupling (at the distance of one lattice constant) is antiferromagnetic for lower relaxations, but becomes ferromagnetic above $15\%$ relaxation. For Ta(110) surface, the exchange couplings for the two nearest neighbors are ferromagnetic for all considered values of relaxations. The weaker nearest-neighbor interaction decreases and the next-nearest-neighbor interaction increases in size with increasing relaxation. 
Also notable is the increasingly antiferromagnetic character of some further (third and fifth) neighbor couplings with increasing relaxation, which will give rise to the formation of a short wavelength spin spiral
along the $[1\overline{1}0]$ direction in Fe/Ta$(110)$, see Sec.~\ref{sec2C}. In particular, this might happen since the strong ferromagnetic coupling between the next-nearest neighbors does not play a role in the formation of the spiral state since it only couples spins along the $[001]$ direction (see coupling $J_{2}$ in Fig.~\ref{fig4}).

The antisymmetric part of the exchange tensors between the Fe atoms is shown in Fig.~\ref{figDij} in terms of  the components of the Dzyaloshinsky--Moriya vectors,
\begin{eqnarray}
D_{ij}^{\alpha}=\frac{1}{2}\sum_{\beta,\gamma}\varepsilon^{\alpha\beta\gamma}J_{ij}^{\beta\gamma}.
\end{eqnarray}
According to the symmetry rules set up by Moriya,\cite{Moriya2} all the Dzyaloshinsky--Moriya vectors lie in the (110) plane. Note that the $x$ and $y$ directions correspond to the $[1\overline{1}0]$ (long) axis and to the $[001]$ (short) axis, respectively. The components of the Dzyaloshinsky--Moriya vectors are only drawn for neighbors with $R_{ij}^{x} \ge 0 $ and $R_{ij}^{y} \ge 0$. The components for the related neighbors can be obtained by symmetry: $(-D_{ij}^{x},D_{ij}^{y})$ for  $(R_{ij}^{x},-R_{ij}^{y})$,
$(D_{ij}^{x},-D_{ij}^{y})$ for  $(-R_{ij}^{x},R_{ij}^{y})$ and $(-D_{ij}^{x},-D_{ij}^{y})$ for  $(-R_{ij}^{x},-R_{ij}^{y})$.
 $D_{ij}^{x}$ is, therefore, only finite between atoms which have a finite distance along the $[001]$ ($y$) direction; for example, the atoms at $\sqrt{2}a$ distance are located along the $[1\overline{1}0]$ ($x$) axis, thus $D_{ij}^{x}=0$.  Similarly, $D_{ij}^{y}$ is only finite if $R_{ij}^{x}\neq 0$. The Dzyaloshinsky-Moriya interactions are comparable in magnitude to the isotropic exchange interactions and they also show oscillating behavior. 

The presence of the Dzyaloshinsky-Moriya interactions may stabilize spin spiral states and the sign of the components of the Dzyaloshinsky-Moriya vectors determines the chirality of the spin spiral. Let $\boldsymbol{q}$ be the wave vector of the spiral, $\boldsymbol{n}$ the normal vector of the monolayer pointing outwards from the substrate, and introduce the vector  
$\boldsymbol{\chi}=\boldsymbol{S}_{i}\times\boldsymbol{S}_{j}$ such that $\boldsymbol{q}(\boldsymbol{R}_{j}-\boldsymbol{R}_{i}) >0 $, where $\boldsymbol{R}_{i}$ and $\boldsymbol{R}_{j}$  are the position vectors of neighboring spins in the lattice. Note that for cycloidal spin spirals the direction of $\boldsymbol{\chi}$ is independent of the choice of the lattice sites $i$ and $j$.  Following Refs.~[\onlinecite{Meckler}] and [\onlinecite{Zimmermann}], a cycloidal spin spiral is called right-rotating when the vectors $\left(\boldsymbol{q},\boldsymbol{\chi},\boldsymbol{n}\right)$ form a right-handed system. If they form a left-handed system, the spin spiral is called left-rotating. With our sign convention and only taking into account the largest Dzyaloshinsky--Moriya interactions in both directions, in the case of W substrate the $D_{ij}^{x}$ component prefers a right-rotating spiral along the $[001]$ direction and the $D_{ij}^{y}$ component prefers a left-rotating spiral along the $[1\overline{1}0]$ direction. This is in agreement with the results in Ref.~[\onlinecite{Zimmermann}] and the chirality of the spin spiral state along the $[001]$ direction in double-layer Fe on W(110).\cite{Meckler} For Ta substrate, the sign of the largest $D_{ij}^{x}$ vector component is flipped compared to the case of W substrate. This means that the Dzyaloshinsky--Moriya interactions prefer left-rotating spirals in an Fe monolayer on Ta(110) along both the $[001]$ and $[1\overline{1}0]$ directions.

\subsection{Ground states obtained from spin dynamics simulations \label{sec2C}}

After obtaining the coupling coefficients from collinear configurations, we performed atomistic spin dynamics simulations to find  the ground states of the systems. These are based on the numerical solution of the Landau-Lifshitz-Gilbert equations,
\begin{equation}
\partial_{t}\boldsymbol{S}_{i}=-\gamma'\boldsymbol{M}_{i}-\alpha \gamma' \boldsymbol{S}_{i}\times\boldsymbol{M}_{i},\label{eqn2}
\end{equation}
with $\gamma'=\frac{1}{1+\alpha^{2}}\frac{ge}{2m}$ the gyromagnetic coefficient ($g$ the g-factor, $e$ the magnitude of charge  and $m$ the mass of the electron) and $\alpha$ the dimensionless Gilbert damping factor. The torque  $\boldsymbol{M}_{i}$ acting on the spin vector $\boldsymbol{S}_{i}$ is defined as
\begin{equation}
\boldsymbol{M}_{i}=\boldsymbol{S}_{i}\times\left(-\frac{1}{m_{i}}\frac{\partial H}{\partial \boldsymbol{S}_{i}}\right),\label{eqn3}
\end{equation}
and $m_{i}$ is the magnitude of the magnetic moment of the atom at site $i$, associated with the spin magnetic moment from the \textit{ab initio} calculations in Sec.~\ref{sec2A}, while $H$ is the spin Hamiltonian in Eq. (\ref{eqn1}).

We also calculated the exchange couplings between the Fe atoms and the atoms in the topmost bulk layer which had the largest induced moment, see Table \ref{table1}. However, we found that including these couplings did not change the ground state considerably, they just give rise to an antiparallel alignment of the induced moments with respect to the neighboring Fe moments. This implies that for the considered systems only the stable Fe moments are relevant to be included into the Hamiltonian (\ref{eqn1}). This feature is essential since the quasiclassical description (\ref{eqn1})-(\ref{eqn2}) is shown to be a reliable description for the rigid moments,\cite{Antropov} but it is probably not valid for the induced moments.

\begin{figure}
\includegraphics[width=\columnwidth]{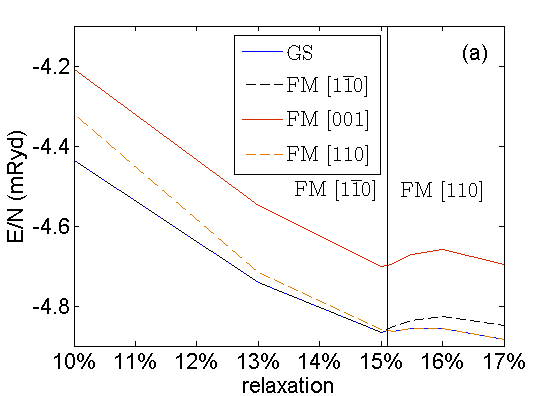}
\includegraphics[width=\columnwidth]{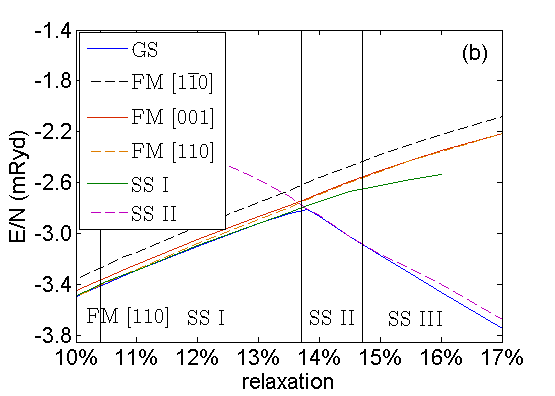}
\caption{\label{fig-gs}(color online) Energies per Fe spin of an Fe monolayer (a) on W$(110)$ and (b) on 
Ta(110) for different magnetic states as a function of the relaxation of the Fe layer obtained from spin dynamics simulations for a system consisting of $N=64\times 64$ atoms with periodic boundary conditions. The energy of the ground state (GS) is highlighted by blue solid line and the types of the ground state magnetic orderings are displayed for the whole range of relaxations. For the explanation of the different spin spiral states (SS I, SS II, SS III) see the text.}
\end{figure}

Starting the spin dynamics simulations from a random initial configuration, the system will generally converge to a metastable equilibrium state, that is to a local energy minimum. However, this configuration may not be the ground state -- the global energy minimum --, therefore the determination of the ground state may require multiple runs. It was found that a random initial state often leads to a spin spiral state, even if it has slightly higher energy than the ferromagnetic state. Furthermore, the obtained equilibrium states may contain skyrmion-like local excitations which are stable with respect to the dynamics of the system, but represent a positive energy correction compared to the ground state.

\begin{figure*}
\includegraphics[width=2.0\columnwidth]{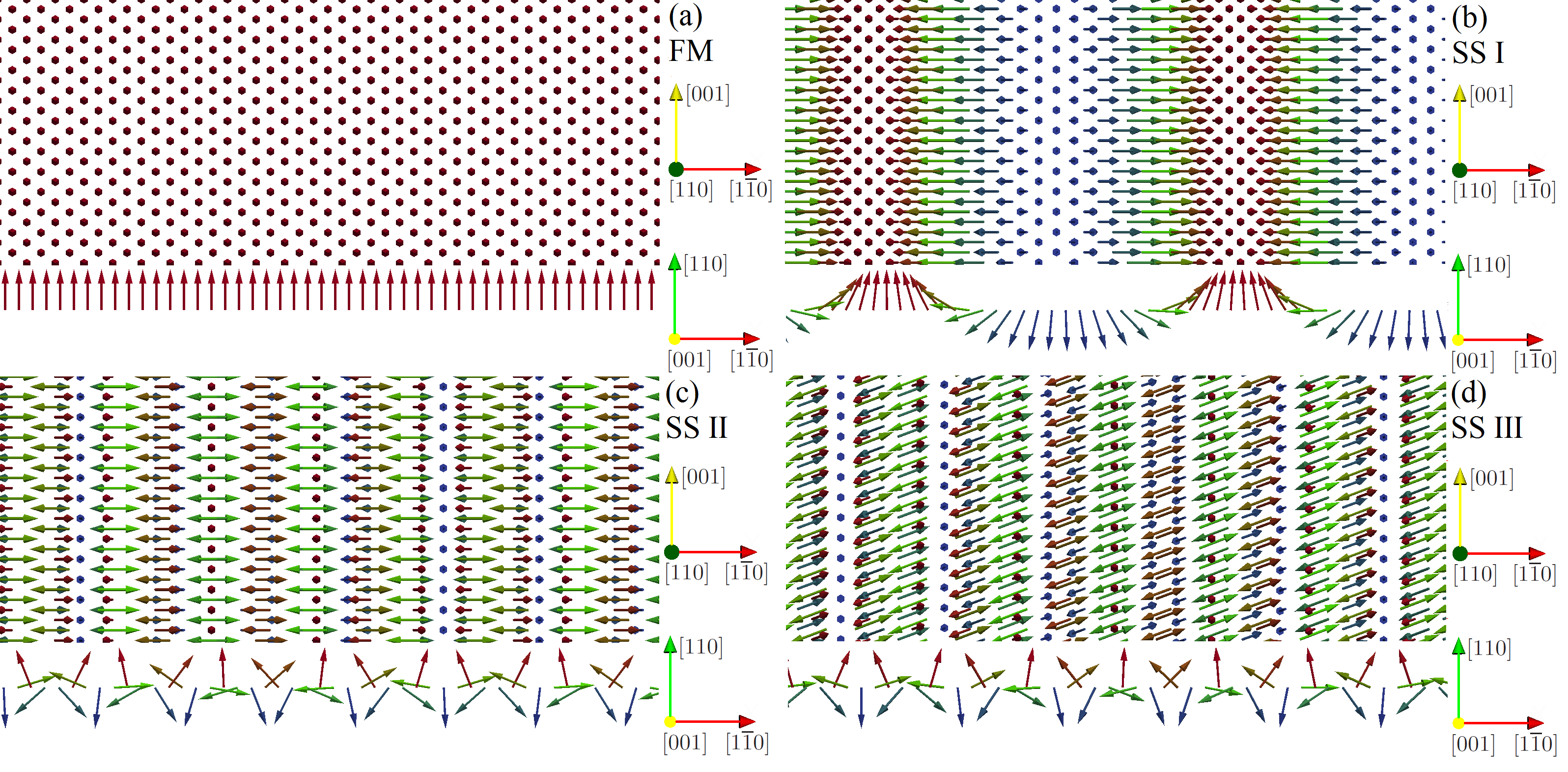}
\caption{\label{fig-spinconf}(color online) Ground state magnetic configurations of an Fe monolayer on Ta$(110)$ for different relaxations: (a) FM state for $10.3\%$, (b) SS~I state for $13.5\%$ ($\lambda \approx 5.98\,\textrm{nm}$), (c) SS~II state for $13.8\%$ ($\lambda  \approx 0.83\,\textrm{nm}$) and (d) SS~III state for $15\%$ ($\lambda \approx 0.81\,\textrm{nm}$). The SS~I and SS~II states differ in the wavelength of the spin spiral, while the SS~II and SS~III states mainly differ in the normal vector of the spiral.}
\end{figure*}

\begin{figure}
\includegraphics[width=\columnwidth]{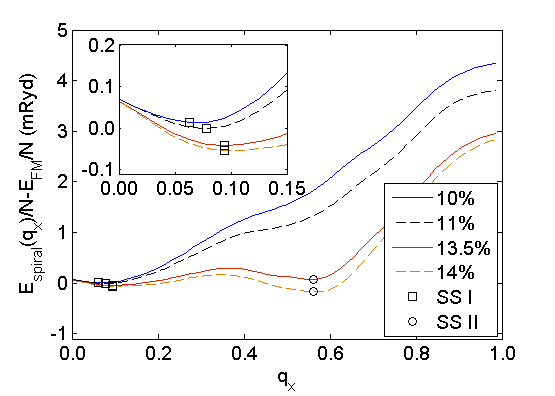}
\caption{(color online) Spin spiral energies per spin relative to the energy of the ferromagnetic state, calculated from the Heisenberg model parameters in the spin spiral configuration Eq. (\ref{eqnspinconf})  for wave vectors along the $[1\overline{1}0]$ direction, $q_x$ (given in units of $\frac{2\pi}{\sqrt{2}a}$). The points at which the spin spiral energies are calculated in Fig.~\ref{fig-gs} are denoted by squares for SS~I and circles for SS~II. The inset shows a magnified view of the range $0 \le q_{x} \le 0.15$.}
\label{figJq}
\end{figure}

The energies obtained from the spin dynamics simulations with the Hamiltonian (\ref{eqn1}) are shown in Fig.~\ref{fig-gs}(a) as a function of the relaxation of the Fe layer in case of W(110). The ground state energy of the system is compared to the energies of the ferromagnetic alignments along the main crystallographic directions $[1\overline{1}0]$, $[001]$ and $[110]$.  The ground state of the Fe monolayer on W$(110)$ was found to be ferromagnetic for all relaxations, however, a reorientation transition occurs at around $15\%$ relaxation of the Fe layer from the in-plane $[1\overline{1}0]$ direction to the out-of-plane $[110]$ direction. The in-plane easy axis at the experimentally observed relaxation value $13\%$ is in agreement with the experiments.\cite{Elmers2} It is worth noting that a double-layer of Fe on W$(110)$ has an out-of-plane easy axis,\cite{Kubetzka,Slezak} similarly to the case here for large relaxation.

In Fig.~\ref{fig-gs}(b) the energies of the ferromagnetic states and also of different spin spiral states are shown for the Fe monolayer on Ta$(110)$. The energies of the cycloidal spiral states SS~I and SS~II were calculated in the homogeneous left-rotating spin spiral configuration,
\begin{eqnarray}
\boldsymbol{S}_{i}=\left(-\sin\left(\boldsymbol{q}\boldsymbol{R}_{i}\right),0,\cos\left(\boldsymbol{q}\boldsymbol{R}_{i}\right)\right), \label{eqnspinconf}
\end{eqnarray}
where the different spin components correspond to the directions $\left(x,y,z\right)=\left([1\overline{1}0],[001],[110]\right)$. 
The normal vector and rotational sense of the spirals chosen in Eq.~(\ref{eqnspinconf}) are consistent with the obtained ground states shown in Fig.~\ref{fig-spinconf}. The spiral energies were calculated for $\boldsymbol{q}$ values in the whole Brillouin zone, but only the $\boldsymbol{q}$ vectors along the $[1\overline{1}0]$ direction, denoted by $q_{x}$, showed complex behavior, see Fig.~\ref{figJq}.  In Fig.~\ref{figJq}, the energy difference between the spin spiral states and the ferromagnetic state along the $[110]$ direction does not go to $0$ as $\boldsymbol{q}\rightarrow\boldsymbol{0}$ due to the anisotropy in the system.

Fig.~\ref{fig-gs}(b) indicates phase transitions at $10.5\%$ relaxation from the ferromagnetic state with out-of-plane easy axis (FM) to the SS~I spin spiral state, at $13.8\%$ relaxation between the SS~I and  SS~II states, and at $14.5\%$ relaxation between the SS~II and SS~III states. All the spiral states have a wave vector parallel to the $[1\overline{1}0]$ direction, and all the spins in the spiral are confined to a plane. For the SS~I and the SS~II states, the spins are located in the $[110]-[1\overline{1}0]$ plane, forming a left-rotating cycloidal spin spiral as in a Mn monolayer on W$(110)$,\cite{Bode} although it is clear from Fig.~\ref{fig-gs}(b) that the $[1\overline{1}0]$ direction is the hard axis since the ferromagnetic state along this direction has the highest energy. The plane of the spiral is thus clearly a consequence of the Dzyaloshinsky--Moriya interaction in the system which prefers spin spiral states oriented perpendicular to the Dzyaloshinsky--Moriya vector. For a spin spiral along the $[1\overline{1}0]$ direction, only the $[001]$ component of the Dzyaloshinsky--Moriya interaction plays a role in the ground state energy, leading to the cycloidal spiral state resembling a N\'{e}el domain wall.

The SS~I state has a small wave number, the value of which increases continuously with increasing relaxation (see the squares in Fig. \ref{figJq}), but jumps to the much larger wave number of the SS~II spin spiral at relaxation 13.8\%. The presence of spin spiral energy minima at different wave vectors and the transition between these minima is a consequence of the frustrated isotropic exchange interactions around these relaxations, see Fig.~\ref{figJij}(b). The SS~III state has similar wave number to the SS~II state, however, the anisotropy is strong enough to rotate the plane of the spiral out from the $[110]-[1\overline{1}0]$ plane, that is the normal vector $[001]$ changes to a general direction in the $[1\overline{1}0]-[001]$ plane. The ground state energies obtained from the spin dynamics simulations in Fig.~\ref{fig-gs}(b) are somewhat lower than the spin spiral energies presented in Fig.~\ref{figJq}, since due to the anisotropy the spiral can gain energy by being deformed with respect to the perfect sinusoidal shape.\cite{Meckler2} This difference is the largest for the SS~III state, but in that case this is also a consequence of the rotation of the normal vector of the spin spiral.

\subsection{Phase transitions at finite temperature using Monte Carlo simulations\label{sec2D}}

\begin{figure*}
\includegraphics[width=\columnwidth]{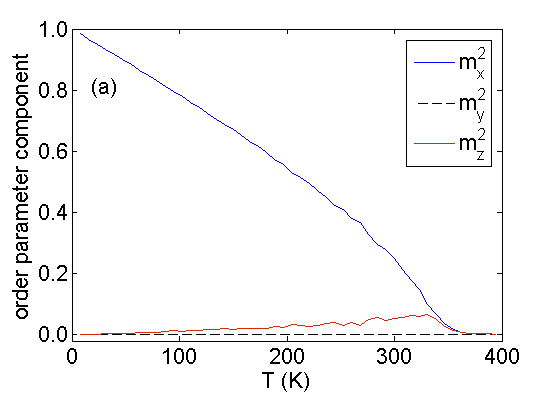}
\includegraphics[width=\columnwidth]{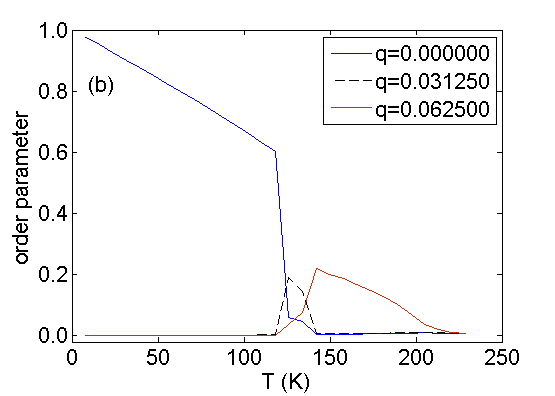}
\includegraphics[width=\columnwidth]{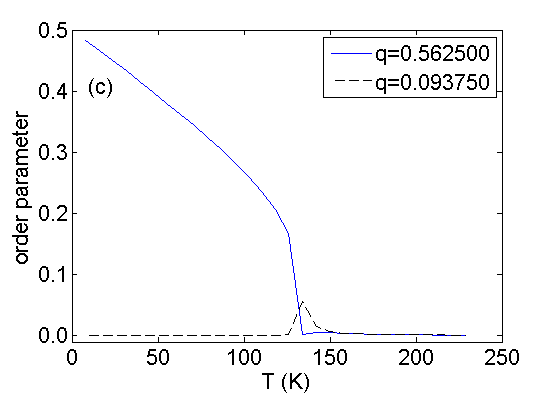}
\includegraphics[width=\columnwidth]{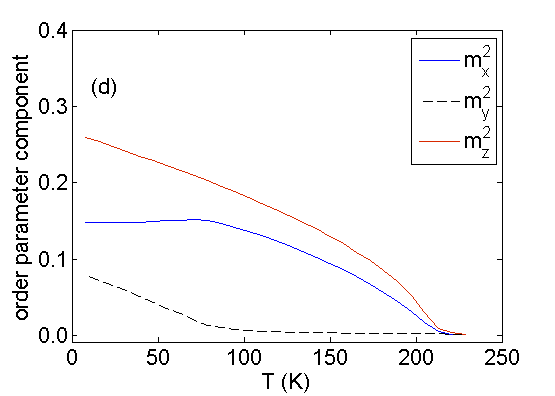}
\caption{(color online) Temperature dependence of the order parameters, Eqs.~(\ref{eqnop})-(\ref{eqnopcomp}), of an Fe monolayer obtained from Monte Carlo simulations for lattice size $N=64\times 64$. The wave number $q$ is given in units of $\frac{2\pi}{\sqrt{2}a}$ and is in all cases parallel the $[1\overline{1}0]$ axis. (a) W(110) surface, $15\%$ relaxation, ferromagnetic order parameter; (b) Ta(110) surface, $10\%$ relaxation; (c) Ta(110) surface, $13.8\%$ relaxation; (d) Ta(110) surface, $15\%$ relaxation, for $q=0.593750$.}
\label{figMC}
\end{figure*}

We examined the phase transitions in the systems also for fixed relaxations as a function of temperature, using classical Monte Carlo simulations with Metropolis dynamics. These phase transitions were expected to occur for relaxation values close to the transition points. The order parameter of the simulations was defined as
\begin{equation}
m^{2}\left(\boldsymbol{q}\right)=\sum_{\alpha=x,y,z}m_{\alpha}^{2}\left(\boldsymbol{q}\right)
,\label{eqnop}
\end{equation}
with
\begin{equation}
m_{\alpha}^{2}\left(\boldsymbol{q}\right)=\left\langle\left|\frac{1}{N}\sum_{i}\textrm{e}^{-\textrm{i}\boldsymbol{q}\boldsymbol{R}_{i}}S_{i}^{\alpha}\right|^{2}\right\rangle,\label{eqnopcomp}
\end{equation}
where $\langle \; \rangle$ denotes thermal average.
As discussed in Sec.~\ref{sec2C}, the shape of the spiral state will differ from a perfect sinusoidal shape due to the anisotropy in the system. Therefore the order parameter for wave vector ${\boldsymbol{q}}$ does not perfectly fit this anharmonic spiral with the same wave vector due to the appearance of higher Fourier harmonics, but still it gives a good approximation to characterize the ordering.\cite{Rocio}

The temperature dependence of the order parameters is shown in Fig.~\ref{figMC}. For the Fe monolayer on W$(110)$, see Fig.~\ref{figMC}(a), no reorientation transition occurred in the system, although the relaxation value of 15\% was close to the transition point. Similarly, no temperature-induced reorientation was found on the other side of the phase boundary, at $15.2\%$ relaxation. The paramagnetic state was reached at $T_{\textrm{c}}\approx 350\,\textrm{K}$, somewhat higher than the experimentally determined critical temperature, $T_{c}\approx 230\,\textrm{K}$.\cite{Elmers} 

In case of the Ta substrate several types of temperature-induced transitions happened between the different ordered phases before reaching the paramagnetic phase, if the chosen relaxation value was close to the phase boundaries shown in Fig.~\ref{fig-gs}(b). The SS~I phase turned out to be the most stable one against thermal fluctuations: systems with ferromagnetic ground state at $10\%$ relaxation or with a SS~II ground state at $13.8\%$ relaxation turned into the SS~I state, in both cases at around $130\,\textrm{K}$, as indicated by a change in the wave number of the order parameter in Fig.~\ref{figMC}(b) and Fig.~\ref{figMC}(c), respectively. Moreover, in case of the FM-SS~I phase transition a continuous  increase of the wave number can be inferred from Fig.~\ref{figMC}(b) above the critical temperature of the phase transition. For the case of a SS~III ground state at $15\%$ relaxation, Fig.~\ref{figMC}(d) shows that the $m_{z}^{2}$ component decreases with the temperature similarly to the order parameter $m^{2}$ in Figs.~\ref{figMC}(b)-(c). However, $m_{x}^{2}$ initially increases with the temperature, which is accompanied by a more pronounced decrease of $m_{y}^{2}$. This indicates that the normal vector of the spin spiral rotates towards the $y=[001]$ axis and at about $80\,\textrm{K}$ a phase transition to the SS~II state occurs. The paramagnetic state was reached at $T_{\textrm{c}}\approx 140-220\,\textrm{K}$ in the case of Ta substrate depending on the relaxation. 

\section{Description of the phase transitions in Fe/Ta$(110)$ based on spin wave expansion\label{sec3}}

\begin{figure}
\includegraphics[width=\columnwidth]{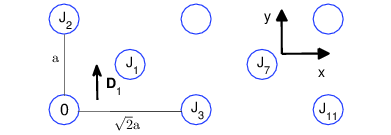}
\caption{(color online) Sketch of the lattice and the model parameters considered in Eq.~(\ref{eqn29}) for an Fe monolayer on Ta$(110)$. $J_{j}$ denote effective exchange couplings between the spin at site $0$ and its neighbors (see text). Equivalent neighbors are formed by mirroring on the $xz$ and $yz$ planes: there are four neighbors of types $1$ and $7$, as well as two neighbors of types $2$, $3$ and $11$. Only the nearest-neighbor Dzyaloshinsky--Moriya vector $\boldsymbol{D}_{1}$ is taken into account, and it transforms as an axial vector.}
\label{fig4}
\end{figure}

In this Section, the temperature-induced phase transitions in the Fe monolayer on Ta(110) surface will be discussed in terms  of spin wave expansion. Keeping the same global coordinate system as in Sec.~\ref{sec2C}, $\left(x,y,z\right)=\left([1\overline{1}0],[001],[110]\right)$, we will use a simplified model Hamiltonian,
\begin{eqnarray}
H=&&\frac{1}{2}\sum_{\substack{i,j \\ (i \ne j)}}
J_{ij}\boldsymbol{S}_{i}\boldsymbol{S}_{j}+\frac{1}{2}\sum_{\substack{i,j \\ (i \ne j)}}\boldsymbol{D}_{ij}\left(\boldsymbol{S}_{i}\times\boldsymbol{S}_{j}\right)\nonumber
\\
&&+\sum_{i}\left[K_{x}\left(S_{i}^{x}\right)^{2}+K_{z}\left(S_{i}^{z}\right)^{2}\right], \label{eqn29}
\end{eqnarray}
where $J_{ij}=J_{ji}$, $\boldsymbol{D}_{ij}=\left(0,D_{ij},0\right)$ with $D_{ij} >0$ for $R^{x}_{ij}>0$ and $D_{ij}=-D_{ji}$, $K_{z} < 0$ and $K_{x} > 0$, that is $z$ is the easy axis and $x$ is the hard axis. We choose the parameters such that the above Hamiltonian reproduces the different phases found in Sec.~\ref{sec2C}. Since the spin spirals have a wave vector parallel to the $x$ axis, only such parameters are relevant which influence the ordering along this direction. These are the effective exchange couplings denoted by $J_{1},J_{2},J_{3},J_{7},J_{11}$  and a Dzyaloshinsky--Moriya vector between the nearest neighbors $\boldsymbol{D}_{1}$ parallel to the $y$ axis (see Fig.~\ref{fig4}). The isotropic couplings are summed up along the $y$ axis: for example, $J_{3}$ represents the coupling between the spin at site $0$ and all the atoms which have the same $x$ coordinate as the third neighbors. This is because the contributions of these Fe-Fe pairs add up in the energy of the spin spirals with wave vectors along the $x$ axis. The anisotropy constants are chosen in agreement with the energies of the ferromagnetic states along the different axes in Fig.~\ref{fig-gs}(b).

Within the spin wave expansion, the energy of the spin system is expanded around a stable equilibrium state using small spin deviations with respect to this state. To lowest order, the Hamiltonian can be written as
\begin{eqnarray}
H_{0}=E_{0}+\sum_{k}\omega_{k}a^{*}_{k}a_{k},\label{eqn17}
\end{eqnarray}
where $E_{0}$ is the energy of the equilibrium state, the $a_{k}$ variables are the classical equivalents of bosonic spin wave annihilation operators and the spin wave energies, $\omega_{k}\ge 0$, stand for the energy corrections due to the spin excitations represented by  $a_{k}$.

For $K_{x}=K_{z}=0$, a homogeneous cycloidal spiral state in the $xz$ plane with wave vector $\boldsymbol{q}_{0}$ along the $x$ axis, $\boldsymbol{S}_{i}=\left(-\sin\left(\boldsymbol{q}_{0}\boldsymbol{R}_{i}\right),0,\cos\left(\boldsymbol{q}_{0}\boldsymbol{R}_{i}\right)\right)$, is either a stable or an unstable equilibrium state of the system. The energy per atom of the spin spiral is given by
\begin{eqnarray}
\frac{E_{0}\left(\boldsymbol{q}_{0}\right)}{N}
=&&\frac{1}{2}J(\boldsymbol{q}_{0})-\frac{1}{2}\textrm{i}D(\boldsymbol{q}_{0}),\label{eqn33}
\end{eqnarray}
with
\begin{eqnarray}
J(\boldsymbol{q})=&&\sum_{j (\ne i)}
J_{ij}\textrm{e}^{-\textrm{i}\boldsymbol{q}\left(\boldsymbol{R}_{i}-\boldsymbol{R}_{j}\right)},
\\
D(\boldsymbol{q})=&&\sum_{j (\ne i)}
D_{ij}\textrm{e}^{-\textrm{i}\boldsymbol{q}\left(\boldsymbol{R}_{i}-\boldsymbol{R}_{j}\right)}.
\end{eqnarray}
The spin wave spectrum around a homogeneous cycloidal spiral state with wave vector $\boldsymbol{q}_{0}$ is given by\cite{Michael,Michael2}
\begin{eqnarray}
\omega_{\boldsymbol{q};\boldsymbol{q}_{0}}=&&\sqrt{C_{+}\left(\boldsymbol{q};\boldsymbol{q}_{0}\right)C_{-}\left(\boldsymbol{q};\boldsymbol{q}_{0}\right)}, \label{eqn32}
\end{eqnarray}
with
\begin{eqnarray}
C_{+}\left(\boldsymbol{q};\boldsymbol{q}_{0}\right)=&&\frac{1}{2}\left[J\left(\boldsymbol{q}-\boldsymbol{q}_{0}\right)+J\left(\boldsymbol{q}+\boldsymbol{q}_{0}\right)\right]\nonumber
\\
&&-\frac{1}{2}\left[\textrm{i}D\left(\boldsymbol{q}+\boldsymbol{q}_{0}\right)-\textrm{i}D\left(\boldsymbol{q}-\boldsymbol{q}_{0}\right)\right]\nonumber
\\
&&-J\left(\boldsymbol{q}_{0}\right)+\textrm{i}D\left(\boldsymbol{q}_{0}\right),\label{eqn30}
\\
C_{-}\left(\boldsymbol{q};\boldsymbol{q}_{0}\right)=&&J\left(\boldsymbol{q}\right)-J\left(\boldsymbol{q}_{0}\right)+\textrm{i}D\left(\boldsymbol{q}_{0}\right),
\end{eqnarray}
where the excitations are indexed with the Fourier transformation wave vectors $\boldsymbol{q}$. The equilibrium state is stable if both $C_{+}\left(\boldsymbol{q};\boldsymbol{q}_{0}\right)$ and $C_{-}\left(\boldsymbol{q};\boldsymbol{q}_{0}\right)$ are non-negative for every $\boldsymbol{q}$, which leads to real and non-negative spin wave frequencies.\cite{Kaplan} The condition $C_{+}\left(\boldsymbol{q};\boldsymbol{q}_{0}\right)\ge 0$ generally holds true if the wave vector $\boldsymbol{q}_{0}$ is close, but not necessarily equal, to the value for which Eq.~(\ref{eqn33}) is minimized. Without Dzyaloshinsky--Moriya interactions, $C_{-}\left(\boldsymbol{q};\boldsymbol{q}_{0}\right)\ge 0$ only holds if $J\left(\boldsymbol{q}_{0}\right)$ is the global minimum of $J\left(\boldsymbol{q}\right)$. However, the presence of the Dzyaloshinsky--Moriya interaction stabilizes several spiral states with different $\boldsymbol{q}_{0}$ values by achieving $C_{-}\left(\boldsymbol{q};\boldsymbol{q}_{0}\right)\ge 0$, even ones which do not minimize Eq.~(\ref{eqn33}). This leads to the appearance of metastable states for which the spin wave expansion (\ref{eqn17}) applies. The presence of the anisotropy may also stabilize these spiral states, either by introducing a hard axis perpendicular to the spiral plane ($K_{y}>0$ in our model, cf. Ref. [\onlinecite{Yosida}]) or by introducing an easy axis in the plane of the spiral ($K_{z}<0$, cf. Ref. [\onlinecite{Miwa}]).

At finite temperatures, the free energy per atom of a system described by the spin-wave Hamiltonian (\ref{eqn17}) can be expressed as
\begin{eqnarray}
\frac{F}{N}=\frac{E_{0}}{N}+\frac{k_{\textrm{B}}T}{N}\sum_{k}\ln\omega_{k}+C\left(T\right),\label{eqn20}
\end{eqnarray}
where $C\left(T\right)$ does not depend on the parameters of the equilibrium state $E_{0}$ and $\omega_{k}$. This expression can describe a transition between two different stable equilibrium states specified by parameters $E_{0},\omega_{k}$ and $E'_{0},\omega'_{k}$. If $E_{0}<E'_{0}$ and the relation $\sum_{k}\ln\omega_{k}>\sum_{k}\ln\omega'_{k}$ applies, then the system will switch from the first state to the second one at the temperature
\begin{eqnarray}
k_{\textrm{B}}T_{\textrm{trans}}=\frac{E'_{0}-E_{0}}{\sum_{k}\ln\omega_{k}-\sum_{k}\ln\omega'_{k}}.\label{eqn23}
\end{eqnarray}
The quantum version of this method was applied in Refs.~[\onlinecite{Yosida}] and [\onlinecite{Miwa}] to describe the transition from a ferromagnetic to a spin spiral state in Dy. It should be noted that this method only gives numerically good transition temperatures if the temperature itself is small, since the spin wave expansion for the free energy (\ref{eqn20}) becomes less accurate as the temperature is increased.

A way of including a perturbative correction in the calculations is by writing the free energy as
\begin{eqnarray}
F=&&E_{0}+\sum_{k}\omega_{k}n_{k}+\frac{1}{2}\sum_{k,k'}P_{kk'}n_{k}n_{k'}\nonumber
\\
&&-k_{\textrm{B}}T\sum_{k}\ln{n_{k}},\label{eqn27}
\end{eqnarray}
where $P_{kk'}$ is a symmetric matrix representing higher order corrections to the energy (\ref{eqn17}) and $n_{k}$ is the occupation number of the spin wave with energy $\omega_{k}$. Minimizing (\ref{eqn27}) with respect to $n_{k}$ leads to self-consistent equations which have real nonnegative solutions only for $T<T_{\textrm{c}}$, giving an estimate of the transition temperature into the paramagnetic phase. This method was originally applied in Ref. [\onlinecite{Bloch}] to find the Curie temperature of a Heisenberg ferromagnet on a simple cubic lattice.

\subsection{The FM-SS I transition\label{sec3A}}

Based on the \textit{ab initio} calculations, we chose different sets of model parameters which are close to the transition points, and employed the spin wave expansion described above to obtain the possible phase transitions as a function of temperature. The calculations were compared to Monte Carlo simulations using the Metropolis algorithm. For $10-11\%$ relaxations the spin spiral energy in Fig.~\ref{figJq} had a single minimum, which we reproduced by choosing a nearest-neighbor ferromagnetic coupling $J_{1}=-2.0\,\textrm{mRyd}$ and a Dzyaloshinsky--Moriya parameter of $D_{1}=0.4\,\textrm{mRyd}$. $K_{z}=-0.22\,\textrm{mRyd}$ was used to move the energy of the ferromagnetic state below the minimum of the spin spiral energy curve. 

As pointed out in Sec.~\ref{sec2C}, the equilibrium spin spiral states of the system are no longer perfect sinusoidal waves, since the spins will prefer the $z$ direction over the $x$ axis. After finding this exact equilibrium state numerically, the spin wave expansion was first performed on the basis of Eq.~(\ref{eqn20}), that is for a system of free spin waves. The results are given in Table \ref{table2} for a lattice size of $N=128\times 64$. The energies of the equilibrium states increase with increasing wave vector, with the ferromagnetic state ($q=0$) being the ground state. Although the size of the lattice influences the allowed wave vectors in the Brillouin zone, the ground state remains ferromagnetic even in the continuum limit\cite{Dzyaloshinsky2,Izyumov} corresponding to an infinite lattice, if the anisotropy is large enough. The free energy correction per spin due to free spin waves ($\sum_{k} \ln \omega_{k} $) decreases when the wave number of the spin spiral increases, leading to the expected transition from the ferromagnetic to the spiral state with increasing temperature. After this transition, the wave number of the equilibrium spin spiral gradually increases. This change is continuous in the continuum limit, therefore the spiral orderings with different wave vectors do not actually represent different phases.

\begin{table}
\begin{ruledtabular}
\begin{tabular}{rrrrr}
$q \left(\frac{2\pi}{\sqrt{2}a_{\textrm{Ta}}}\right)$ &     $\lambda$ (nm) &    $\frac{E_{0}}{N}$ (mRyd)&          $\frac{F_{\textrm{SW}}}{Nk_{\textrm{B}}T}$ &         $T_{\textrm{trans}}^{\textrm{free}}$ (K)\\
         $0.000000$ &  $\infty$&    $-4.2200$ &     $1.9630$ &     $0.0$ \\
  $0.015625$ & $29.88$&  $-4.2176$ &       $1.9531$ &  $38.2$ \\
   $0.031250$ & $14.94$&   $-4.2152$ &     $1.9438$ &  $40.7$ \\
  $0.046875$ & $9.96$&   $-4.2124$ &     $1.9353$ &   $51.9$ \\
    $0.062500$ & $7.47$&   $-4.2072$ &     $1.9282$ &   $115.5$ \\
  $0.078125$ &  $5.98$&  $-4.1966$ &     $1.9215$ &  $249.4$ \\
   $0.093750$ &  $4.98$&  $-4.1786$ &     $1.9138$ &   $368.6$ \\
\end{tabular}
\end{ruledtabular}
\caption{Energy ($E_0/N$), free energy correction ($F_{\textrm{SW}}/Nk_{\textrm{B}}T=\sum_{k}\ln\omega_{k}/N$) per spin and transition temperature ($T_{\rm trans}^{\textrm{free}}$) as defined  in Eq.~(\ref{eqn23}) for different wave numbers ($q$) and corresponding wavelengths ($\lambda$) for the FM-SS~I transition, calculated for a lattice of $N=128\times 64$ atoms with periodic boundary conditions.}  \label{table2}
\end{table}

\begin{table}
\begin{ruledtabular}
\begin{tabular}{rrrrr}
$q \left(\frac{2\pi}{\sqrt{2}a_{\textrm{Ta}}}\right)$ &   $\lambda$ (nm) &      $T_{\textrm{trans}}$ (K) &          $T^{\rm free}_{\textrm{trans}}$ (K) &         $T_{\textrm{c}}$ (K) \\
         $0.000000$ &  $\infty$&    $0.0$ &     $0.0$ &     $201.8$ \\
   $0.031250$ &  $14.94$&  $49.4$ &     $38.4$ &  $223.9$ \\
    $0.062500$ &  $7.47$&  $88.5$ &     $91.1$ &   $261.7$ \\
\end{tabular}
\end{ruledtabular}
\caption{Transition temperatures for different wave numbers ($q$) and corresponding wavelengths ($\lambda$), calculated for a lattice size of $N=64\times 32$ with periodic boundary conditions. $T_{\textrm{trans}}$ and $T^{\rm free}_{\textrm{trans}}$ indicate the temperature where the SS~I spiral with the given wave number becomes the global minimum of the free energy derived from the perturbation theory, Eq. (\ref{eqn27}), and for free spin waves, Eq. (\ref{eqn20}), respectively. $T_{\textrm{c}}$ is the temperature where the state becomes unstable according to perturbation theory.}  \label{table3}
\end{table}

\begin{figure}
\includegraphics[width=\columnwidth]{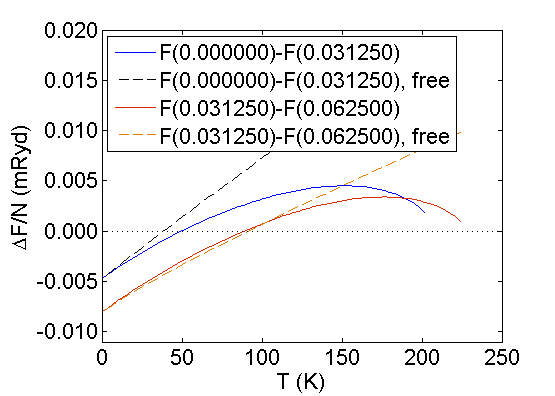}
\caption{(color online) Temperature dependence of free energy differences $\Delta F/N$ between different states, including the ferromagnetic state and spin spirals at different wave numbers. The line at $\Delta F/N=0$ is a guide to the eye, identifying the transition temperatures. The differences obtained with perturbation theory, Eq. (\ref{eqn27}), are compared to the linear functions of the free spin wave theory, Eq. (\ref{eqn20}), for a lattice size of $N=64 \times 32$. The wave numbers are given in units of $\frac{2\pi}{\sqrt{2}a}$.}
\label{fig-fdiff}
\end{figure}

Including perturbation corrections in the calculations on the basis of Eq.~(\ref{eqn27}) makes it possible to give an approximation for $T_{\textrm{c}}$, where the equilibrium state loses its stability and becomes paramagnetic. This gives an upper bound for the transition temperatures, $T_{\textrm{trans}}$. The results are summarized in Table~\ref{table3}. It is worth noting that although the ferromagnetic state remains metastable for a wide temperature range in the SS~I phase, there is a temperature region where only the spiral state is stable and the ferromagnetic state becomes paramagnetic, in agreement with the prediction of Ref. [\onlinecite{Yosida}]. Including the perturbative correction also modifies the transition temperature $T_{\textrm{trans}}$ compared to the non-interacting case. The transition temperature from the ferromagnetic state to the first spiral state is significantly increased for the interacting case, as can be inferred from Fig.~\ref{fig-fdiff} and Table~\ref{table3}. Interestingly, the transition temperature between the spin spiral states with different wave vectors is hardly affected by the perturbation correction. Note that the $T_{\textrm{trans}}^{\textrm{free}}$ transition temperatures are slightly different in Tables \ref{table2} and \ref{table3} because of the different lattice sizes used in the calculations. The reason for this is that the lattice size influences not only the allowed $q$ values, but also the spin wave energies.

\begin{figure}
\begin{center}
\includegraphics[width=\columnwidth]{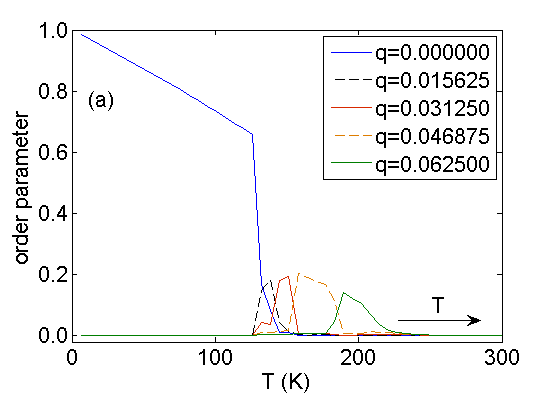}
\includegraphics[width=\columnwidth]{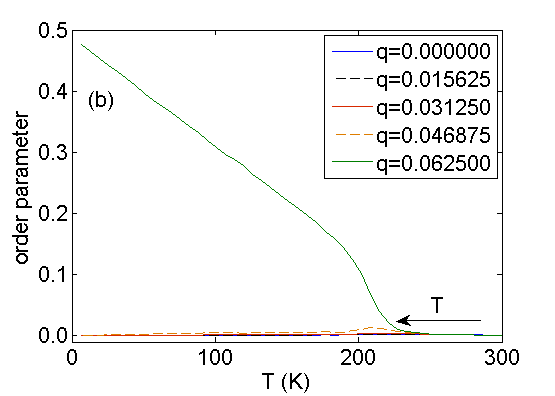}
\caption{(color online) The order parameter defined in Eq.~(\ref{eqnop}) for different wave numbers obtained from Monte Carlo simulations as a function of (a) increasing and (b) decreasing temperature, describing the FM-SS I transition, for a lattice size of $N=128\times 64$. The wave numbers are given in units of $\frac{2\pi}{\sqrt{2}a}$.}
\label{fig-mc}
\end{center}
\end{figure}

Fig.~\ref{fig-mc} shows the results of Monte Carlo simulations for the same model system. As clear from Fig.~\ref{fig-mc}(a), the simulation results are in good agreement with the spin wave calculations: starting from a ferromagnetic ground state, the system will turn into a spiral state with gradually increasing wave vector until the temperature becomes high enough to remove all kinds of magnetic order from the system. The values for $k_{\textrm{B}}T_{\textrm{trans}}$ are somewhat inaccurate (compare Tables \ref{table2}-\ref{table3} with Fig.~\ref{fig-mc}), mainly because the transitions apparently show hysteresis. The lower wave vector states will remain metastable at higher temperatures than the point where the free energy minimum moves to a different wave vector (see Fig.~\ref{fig-fdiff}). This is even more pronounced in Fig.~\ref{fig-mc}(b), where the simulation was performed for decreasing temperature, starting from a random initial state. Although the $q=0.062500$ state is not the ground state, the system freezes into this metastable state in this case. On the other hand, the transition point to the paramagnetic state $T_{\textrm{c}}$ is well approximated by the perturbation theory: for the $q=0.062500$ spiral state, it predicts $T_{\textrm{c}}=261.7\,\textrm{K}$, while the critical temperature from the simulation is around $220\,\textrm{K}$. For comparison, the random phase approximation\cite{Tyablikov} gives $T_{\textrm{c}}=271.1\,\textrm{K}$ for the critical temperature of the ferromagnetic state. The same kind of transition was obtained using the \textit{ab initio} coupling coefficients instead of the model parameters, compare Fig.~\ref{figMC}(b) with Fig. \ref{fig-mc}(a).

\subsection{The SS II-SS I transition}

The SS II-SS I transition can be examined using the same methods as in the previous case. The main difference is that the energy of the spin spiral  must have two different minima, both corresponding to spiral orderings, that is $\boldsymbol{q}_{1},  \boldsymbol{q}_{2} \ne \boldsymbol{0}$ (see Fig.~\ref{figJq}). This requires at least four different coupling coefficients in the spin model (\ref{eqn29}) along the $x$ axis, illustrated in Fig.~\ref{fig4}. For the model calculations we chose $J_{1}=-2.0\,\textrm{mRyd}$, $J_{3}=2.58\,\textrm{mRyd}$, $J_{7}=-1.0\,\textrm{mRyd}$ and $J_{11}=0.8\,\textrm{mRyd}$, which could reproduce the shape of the curves in Fig.~\ref{figJq}, with a slightly lower minimum at high wave number $q=0.593750$ and a somewhat higher one at $q=0.156250$. We also considered the same Dzyaloshinsky--Moriya interaction between the nearest neighbors as in the previous case, $D_{1}=0.4\,\textrm{mRyd}$, since this is necessary to stabilize both spiral states at zero temperature, see Eq. (\ref{eqn32}) and the subsequent discussion. We omitted the anisotropy terms needed to make the ferromagnetic state energetically favorable in Sec.~\ref{sec3A}, since they are irrelevant for the current discussion. The energies and free energy corrections are given in Table~\ref{table4}, for lattice size of $N=128\times 64$.

\begin{table}
\begin{ruledtabular}
\begin{tabular}{rrrrr}
$q \left(\frac{2\pi}{\sqrt{2}a_{\textrm{Ta}}}\right)$ &     $\lambda$ (nm) &         $\frac{E_{0}}{N}$ (mRyd) &          $\frac{F_{\textrm{SW}}}{Nk_{\textrm{B}}T}$ &         $T_{\rm trans}^{\textrm{free}}$ (K) \\
         $0.593750$ &   $0.79$&   $-2.9894$ &     $1.4779$ &     $0.0$ \\
  $0.156250$ & $2.99$&  $-2.9736$ &       $1.4168$ &  $40.8$ \\
\end{tabular}
\end{ruledtabular}
\caption{Energy ($E_0/N$), free energy correction ($F_{\textrm{SW}}/Nk_{\textrm{B}}T=\sum_{k}\ln\omega_{k}/N$) per spin and transition temperature ($T_{\rm trans}^{\textrm{free}}$) values as in Table~\ref{table2} for different wave numbers (or wavelengths) for the SS~II-SS~I transition, for a lattice size of $N=128\times 64$ with periodic boundary conditions.}  \label{table4}
\end{table}

\begin{figure}
\begin{center}
\includegraphics[width=\columnwidth]{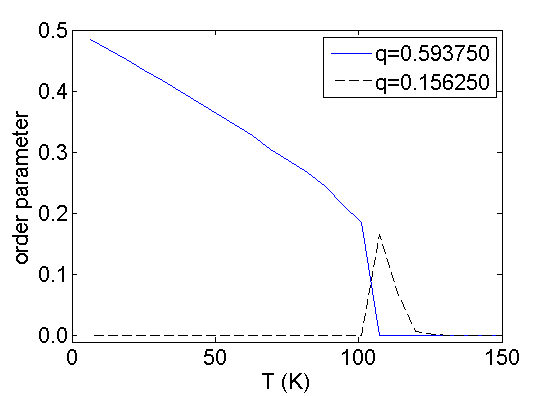}
\caption{The order parameter for different wave numbers as a function of temperature obtained from Monte Carlo simulations for the SS~II-SS~I transition, for a lattice size of $N=128\times 64$. The wave numbers are given in units of $\frac{2\pi}{\sqrt{2}a}$.}
\label{fig9}
\end{center}
\end{figure}

The spin wave calculations indicate that starting from a high wave vector ground state, the system may indeed switch to a low wave vector ordering. This is in agreement with the Monte Carlo simulation results with the same parameter set, shown in Fig.~\ref{fig9}, as well as simulations performed with the \textit{ab initio} coupling coefficients, see Fig.~\ref{figMC}(c). The spin wave expansion again underestimates the transition temperature as in the case of the FM-SS~I transition.

\subsection{The SS~III-SS~II transition}

The third type of transition found in the Fe monolayer on Ta$(110)$ surface corresponds to the case when the wave vector of the spiral remains fixed, but the spiral normal vector rotates from the $y$ axis (the cycloidal state) towards a direction in the $xy$ plane. For modelling this transition we supposed that the wave vector $\boldsymbol{q}_{0}$ of the spiral state is determined by the isotropic exchange couplings, while the Dzyaloshinsky--Moriya interaction and the anisotropy terms were taken into account as a perturbation. For the anisotropy we chose $K_{x}>0$ and $K_{z}=0$, since \textit{ab initio} calculations indicated that at 15\% relaxation the ferromagnetic states along the $y$ and $z$ axes have almost the same energy, while the $x$ axis is a hard axis (see Fig.~\ref{fig-gs}(b)). The angle between the $xz$ plane and the plane of the spin spiral will be denoted by $\varphi$. In this case, the energy contribution per spin from the Dzyaloshinsky--Moriya interaction and the anisotropy terms can be expressed as
\begin{eqnarray}
\frac{\Delta E}{N}=-\frac{1}{2}\textrm{i}D\left(\boldsymbol{q}_{0}\right)\cos\varphi+\frac{1}{2}K_{x}\cos^{2}\varphi. \label{eqn39}
\end{eqnarray}

Differentiating (\ref{eqn39}) with respect to $\varphi$ leads to the stationary points

\begin{eqnarray}
\sin\varphi^{(1)}=&&0,
\\
\cos\varphi^{(2)}=&&\frac{\textrm{i}D\left(\boldsymbol{q}_{0}\right)}{2K_{x}}. \label{eqn41}
\end{eqnarray}
Substituting the solutions into (\ref{eqn39}) gives
\begin{eqnarray}
\frac{\Delta E^{(1)}}{N}=&&\mp\frac{1}{2}\textrm{i}D\left(\boldsymbol{q}_{0}\right)+\frac{1}{2}K_{x}, \label{e1}
\\
\frac{\Delta E^{(2)}}{N}=&&-\frac{\left(\textrm{i}D\left(\boldsymbol{q}_{0}\right)\right)^{2}}{8K_{x}},  \label{e2}
\end{eqnarray}
implying that whenever the second stationary point exists,
\begin{eqnarray}
\left|\frac{\textrm{i}D\left(\boldsymbol{q}_{0}\right)}{2K_{x}}\right|<1, \label{cond}
\end{eqnarray}
it will correspond to the energy minimum. This describes the rotation of the spiral normal vector away from the $y$ axis when the Dzyaloshinsky--Moriya interaction is weak compared to the anisotropy. Calculating the spin wave spectrum reveals that only one of the states is stable for any value of $D$ and $K_{x}$, therefore the spin wave expansion is not suitable for describing this type of transition.

For the present simulations the exchange parameters $J_{1}=-2.0\,\textrm{mRyd}$, $J_{3}=3.0\,\textrm{mRyd}$ and $J_{7}=-1.0\,\textrm{mRyd}$ were chosen which lead to a spin spiral along the $x$ axis with a wave number $q=0.546875$ ($\lambda=0.85\,\textrm{nm}$). We took $D_{1}=0.05\,\textrm{mRyd}$ between the nearest neighbors and $K_{x}=0.2\,\textrm{mRyd}$, and found that these values did not influence the shape of the spiral considerably, but confined the spins to a plane with a normal vector lying in the $xy$ plane, as shown in Fig.~\ref{fig-spinconf}(d). We also used a ferromagnetic coupling between the neighbors in the $y$ direction, $J_{2}=-2.0\,\textrm{mRyd}$, which does not influence the spiral state but removes the possible domain walls from the system along the $y$ axis. These domain walls occur because Eq.~(\ref{eqn41}) has two solutions $\pm\varphi^{(2)}$ with the same energy, therefore if the spins are weakly coupled along the $y$ direction, $\varphi^{(2)}$ and $-\varphi^{(2)}$ domains may be simultaneously present in the system. 

\begin{figure}
\begin{center}
\includegraphics[width=\columnwidth]{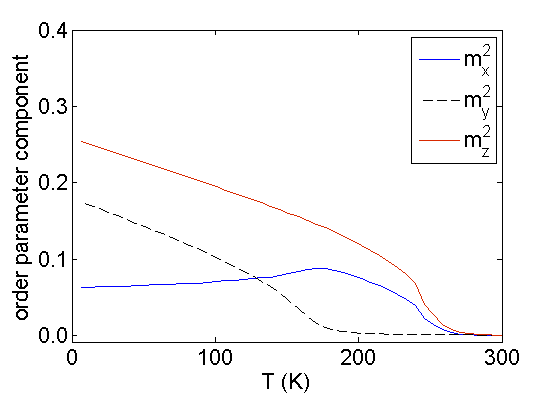}
\vskip -10pt
\caption{(color online) Different components of the order parameter Eq.~(\ref{eqnopcomp}) as a function of temperature, for the SS~III-SS~II transition. The wave number of the spin spiral was $q=0.546875 \frac{2\pi}{\sqrt{2}a}$ and a lattice size of $N=128\times 64$ was used.}
\label{fig10}
\end{center}
\end{figure}

The SS~III-SS~II transition is shown in Fig.~\ref{fig10}, in agreement with the simulations performed with \textit{ab initio} coupling coefficients, see Fig.~\ref{figMC}(d). By increasing the temperature, the plane of the normal vector of the spiral rotates towards the $y$ axis, which is the one preferred by the Dzyaloshinsky--Moriya interaction over the $x$ direction preferred by the anisotropy. This indicates that with increasing temperature the magnitude of the effective Dzyaloshinsky--Moriya contribution to the free energy decreases slower than the anisotropy contribution.

\section{Summary and conclusions}

We examined the phase diagram of an Fe monolayer on the $(110)$ surfaces of W and Ta as a function of the relaxation of the Fe layer and the temperature. We used the relativistic screened Korringa-Kohn-Rostoker method to determine the single-particle potential of the systems within the local density approximation of density functional theory. In terms of the relativistic torque method, we calculated the tensorial coupling coefficients which appear in the generalized Heisenberg model describing the spin system, Eq.~(\ref{eqn1}). Based on this spin model, we determined the magnetic ground state from spin dynamics simulations, and we performed Monte Carlo simulations to explore the magnetic phase transitions at finite temperature.

In case of W substrate the obtained magnetic moments and ground states were in good agreement with previous calculations\cite{Qian2,Qian3,Huang} and with experiments.\cite{Elmers2} The ground state was ferromagnetic with an easy axis along the $[1\overline{1}0]$ direction for relaxations smaller than $15\%$, including the experimentally and theoretically determined relaxation values around $12\%-13\%$. For larger relaxations, the system remained ferromagnetic, but the easy axis turned into the out-of-plane $[110]$ direction. For fixed relaxations, we found no thermally induced transition between these two states.

In case of Ta substrate four different phases were identified in the considered relaxation range, see Fig.~\ref{fig-gs}(b), with transitions occurring at $10.5\%$, $13.8\%$ and $14.5\%$ relaxation values.  At low relaxations the ground state was ferromagnetic with an easy axis along $[110]$. The next two phases, denoted by SS~I and SS~II, correspond to cycloidal spin spirals with wave vectors along the $[1\overline{1}0]$ direction and normal vector along the $[001]$ axis, the SS~II state having a significantly larger wave number. The SS~II and SS~III spin spirals had similar wave vectors but the normal vector of the spiral left the $[001]$ axis in the SS~III state.

Choosing the relaxation value close to one of the transition points, different types of transitions were obtained between these states at finite temperature. These possible phase transitions were described theoretically using spin wave expansion and compared to Monte Carlo simulations performed on model systems. We found that starting from a ferromagnetic ground state, the system may turn into a spin spiral state at finite temperature before becoming paramagnetic. Although the appearance of the spin spiral state as a consequence of the Dzyaloshinsky--Moriya interaction is a well-known effect in two-dimensional systems such as Mn monolayer\cite{Bode} or Fe double-layer\cite{Meckler} on W$(110)$, there was no such transition observed as a function of temperature. However, \textit{ab initio} calculations\cite{Heide,Zimmermann,Bergqvist} indicated a ferromagnetic ground state for Fe double-layer on W$(110)$, suggesting that this system is probably very close to such a ferromagnetic-spin spiral transition. We have also shown that the high wave vector SS~II state may turn into the low wave vector SS~I spiral by increasing the temperature, while in the case of the SS~III-SS~II transition the normal vector of the spin spiral rotated from a general in-plane direction towards the $[001]$ direction. 

For all three phase transitions, the simulations performed on model systems and using the \textit{ab initio} coupling coefficients gave results which were in agreement with the predictions based on spin wave expansion. Compared to the Monte Carlo simulations, the spin wave expansion gave good approximations for the temperature $T_{\textrm{c}}$ where any magnetic order disappears and somewhat underestimated the transition temperature $T_{\textrm{trans}}$ between the ordered states. The latter difference is also a consequence of the metastability of the states, indicating that conventional Monte Carlo simulations are not well-suited for finding the actual transition temperature.

Given the wide variety of possible ground states in a relatively narrow range of relaxations, our present work might motivate experiments to determine the actual magnetic ground state of an Fe monolayer on Ta$(110)$. It may even be possible to find one of the thermally induced transitions described here. On the other hand, the spin wave expansion method may also be applied for the finite-temperature description of other stable equilibrium configurations such as the skyrmion lattice structure found in ultrathin magnetic films.\cite{Heinze,Romming}

\begin{acknowledgments}
The authors thank Eszter Simon, L\'{a}szl\'{o} Ujfalusi and Bernd Zimmermann for enlightening discussions. Financial support was provided by the Hungarian Scientific Research Fund under project No. K84078 and by the European Union under FP7 Contract No. NMP3-SL-2012-281043 FEMTOSPIN. The work of  LS and IAS was also supported by the project
T\'{A}MOP-4.2.2.A-11/1/KONV-2012-0036 co-financed by the European Union and the European Social Fund.
\end{acknowledgments}

\end{document}